\documentstyle[preprint,aps]{revtex}
\begin{document}
\draft
%\preprint{\vbox{\hbox{IMSC/00-xx-xx}\smallskip\hbox{cond-mat/0107270}}}
\title{The Hydrodynamical Limit of Quantum Hall system}
\author{D. Sreedhar Babu$^*$, R. Shankar}
\address{The Institute of Mathematical Sciences,\\
C.I.T Campus, Chennai - 600 113, INDIA.}
\author{M. Sivakumar}
\address{School of Physics, University of Hyderabad,\\
Hyderabad, INDIA}
\footnotetext {$^*$ Currently at,\\ Theoretical Physics Group,
Tata Institute of Fundamental Research, Bombay 400 005. }  
\maketitle
\begin{abstract} 

We study the current algebra of FQHE systems in the hydrodynamical limit
of small amplitude, long-wavelength fluctuations. We show that the
algebra simplifies considerably in this limit. The hamiltonian is
expressed in a current-current form and the operators creating
inter-Landau level and lowest Landau level collective excitations are
identified.
\end{abstract}

\newpage

\section{Introduction}

Theoretical insights into the physics of Fractional Quantum Hall Effect 
(FQHE) systems was initially obtained by the study of the pioneering 
variational wavefunctions proposed by Laughlin \cite{laugh1} and by the 
exact diagonalisation of small systems. The concept of composite fermions, 
put forward by Jain \cite{jain1} gave further insight into the physics and
a unified picture of all the observed integer and fractional values of
the filling factor, $\nu$. The composite fermions can be modelled as 
fermions interacting with a Chern-Simons gauge field. The observed
fractions can be obtained as mean field solutions of this theory.

While attempts to treat the fluctuations about the mean field state have
had some success \cite{hlretc}, there is really no systematic expansion
as yet. Recently, G.Murthy and R. Shankar \cite{gmrs1} have proposed a 
Hamiltonian formalism using which they develop an analytic approximation 
scheme in the long wavelength limit. Based on these results, R. Shankar 
\cite{rs1} has formulated a calculational prescription that goes beyond the
long wavelength limit. This prescription has been sucessful in
analytically computing the excitation spectrum of the system with a fair
degree of accuracy.

Our work is motivated by the success of the abovementioned works. The
approximations used in reference \cite{gmrs1} are valid for
longwavelength, small amplitude density fluctuations, namely in the
hydrodynamic limit. In this limit, current algebra techniques have been 
used with success in other systems of interacting fermions. These span a huge
range in energy and length scales. From 1d electrons in solids (energies
$\sim$ meV and wavelengths $\sim$ A) to interacting quarks (energies $\sim$ 
MeV and wavelengths $\sim$ fm). In these cases, the theory can be
written in terms of the hydrodynamic variables, the current density and
charge density operators. The spectrum of collective excitations 
(particle-hole pairs) is described by small amplitude harmonic
oscillations of the fluid and the charged, fermionic quasiparticles manifest 
as topological solitons. Such a picture is physically well established in FQHE
systems. The ground state is a an incompressible liquid, the low energy
collective excitations are small amplitude density fluctuations and the
charged quasiparticles (in this case fractionally charged anyons) are vortices. 
Thus current algebra techniques should be useful here also and could possibly 
be used to develop a systematic long-wavelength expansion. 

This work is our first step in this direction. We first develop a fully
field theoretic version of the formalism in reference \cite{gmrs1}.
In section \ref{cf}, we begin with the system of electrons and
construct a unitary flux attatchment operator that transforms electrons
into flux carrying composite fermions coupled to a Chern-Simons gauge 
field. Section \ref{mfs} constructs the "mean field" states and shows
that they correspond to the (unprojected) Jain states. We then derive 
the current algebra of the composite fermions and its hydrodynamic limit,
in section \ref{ca}. In this limit, section \ref{th} expresses the hamiltonian
in terms of the electron current and charge densities. The collective
excitations and the lowest Landau level (LLL) projection is discussed in
section \ref{ce}. We summarise our results in section \ref{sum}.

\section{Composite Fermions}
\label{cf}

\subsection{The electronic system}

We start with the system of interacting electrons in a magnetic field.
The electron anihilation and creation field operators, $C(x)$ and 
$C^\dagger(x)$ satisfy the anticommutation relations,
\begin{equation}
\{C(x), ~C(y)^{\dagger}\} = \delta^2(x-y)
\end{equation}
The Hamiltonian of the system is $ H= H_0 + H_{int}$ where,
\begin{eqnarray}
H_0=\int d^2 x C(x)^{\dagger}\frac{1}{2} (-i \frac{\partial}{\partial
x^i}+ \frac{1}{2}\epsilon_{ij}x^j)^2 C(x)\label{ham0}\\
H_{int}=\frac{e^2}{2\kappa}\int d^2 x d^2 y U(x-y)C(x)^{\dagger}C(y)^{\dagger}  
C(y)C(x)\label{ham.int}
\end{eqnarray}

We have expressed all lengths in units of $l_c~( \frac{1}{l_c^2}
=\frac{eB}{\hbar c} $) and all energy scales in units of$h\omega_c =
\frac{eB}{m^* c}$. In these units we have, 

\begin{eqnarray}    
\label{units}
A_i=-\frac{1}{2} \epsilon_{ij}x^j\\
\nonumber
B=1 ~{\rm and}~ {\bar\rho} =\frac{\nu}{2\pi}
\end{eqnarray}
\label{cden}
The charge density and current operators are given by
\begin{eqnarray}
\rho(x)=C(x)^{\dagger}C(x)\\
\nonumber
J_i(x)=\frac{1}{2i}(C^{\dagger}{\partial}_iC-{\partial}_iC^{\dagger}C)-A_i(x)
{\rho}
\end{eqnarray}

\subsection{The extended Hilbert Space}
We now introduce the Chern-Simons gauge fields, $a^\prime_i(x)$, obeying the 
commutation relations,

\begin{equation}
\label{apcom}
[a^\prime_(x), a^\prime_j(y)]=i4{\pi k}{\epsilon}_{ij}{\delta}^2(x-y),
~~~(k= 1,2,3...)
\end{equation}
equivalently,  $a^\prime(x)\equiv (a^\prime_1+ia^\prime_2)/\sqrt{2\pi k}$ 
obey the oscillator algebra
\begin{equation} 
[a^\prime(x), a^\prime(y)^{\dagger}]={\delta}^2(x-y)
\end{equation}

We denote the Hilbert space of the Chern-Simons fields as 
${\cal H}^\prime_{CS}$ and
call the electronic Hilbert space ${\cal H}_{el}$. We then define an extended
Hilbert space, 
\begin{equation}
{\cal H}_{ex}\equiv{\cal H}_{el}\otimes{\cal H}^\prime_{CS}
\end{equation}
A subspace ${\cal H}_{phy}$ is defined as the set of states satisfying 
the constraint,
\begin{equation}
\frac{1}{4\pi k}\epsilon_{ij}\partial_ia_j|\psi>_{phy} = 0
\end{equation}
It can be shown that a unique solution exists to this constraint in 
${\cal H}^\prime_{CS}$. Therefore we trivially have,
\begin{equation}
{\cal H}_{phy} = {\cal H}_{el}
\end{equation}

\subsection{Flux attatchment}

	Next we construct an unitary operator, which attatches  
2k units of Chern-Simons flux to the electrons. Namely, an unitary
operator that transforms electrons into composite fermions. This
operator, $V$, is given by,
\begin{equation}
\label{vop}
 V= e^{i({\int d^2 x d^2y \frac{-k}{2}\rho(x)\theta(x-y)\rho(y) +\int\rho(x)
d^2x d^2y \epsilon_{ij}\alpha_i^v (x-y) a_j(y)})}
\end{equation}
where,
\begin{eqnarray}
\alpha^v_i &=& \epsilon_{ij}\partial_j \Theta(x-y)\\
\epsilon_{ij}\partial_i\alpha_j^v (x-y) &=& 2\pi\delta^2(x-y)\\ 
\end{eqnarray}
It can be shown that,
\begin{eqnarray}
\label{vcv}
 V C(x) V^{\dagger} &=& e^{i2k(\Omega_L(x)-\Phi(x))C(x)}\\
\nonumber
		    &\equiv&\Psi(x)\\
\label{vav}
 V a^\prime_i(x) V^{\dagger}&=&a^\prime_i(x) + 2k\int d^2y (\alpha)^v(x-y)
\rho(y)\\
\nonumber
		     &\equiv& a_i(x)
\end{eqnarray}  
where,
\begin{eqnarray}
\label{omdef}
\Omega_L(x) &\equiv& \frac{1}{4\pi k}{\int}_y\epsilon_{ij}
\alpha_{i}^v(x-y)a_j(y)\\
\Phi(x) &\equiv& {\int}_y\theta(x-y)\rho(y)
\end{eqnarray}  

Since $\Psi$ and $a_i$ are unitary transforms of $C$ and $a^\prime_i$,
they satisfy the same commutation relations,
\begin{eqnarray}
\label{psicom}
\{\Psi (x), \Psi^{\dagger}(y)\} = \delta^{2} (x-y)\\
\label{acom}
\left [ a_i (x), a_j (y) \right ] = i 4\pi k \delta^{2}(x-y).
\end{eqnarray}
Equations (\ref{omdef}) and (\ref{acom}) imply that  $\Omega_L$ is 
congugate to the Chern-Simons magnetic field i.e, 
\begin{equation}  
[\Omega_L(x), \epsilon_{ij}\partial_i a_j(y)]=i2\pi\delta^2(x-y)
\end{equation}
Therefore, $e^{i2k\Omega}$ creates a vortex carrying 2k units of flux at x. 
$e^{-2ik\Phi(x)}$ gives the corresponding Ahronov-Bohm Phase. Thus 
$\Psi^\dagger(x)$ creates an electron and a vortex carrying 2k units of flux 
at $x$, namely it creates a composite fermion.

Thus if we define $V{\cal H}^\prime_{CS}V^\dagger \equiv {\cal H}_{CS}$
and $V{\cal H}_{el}V^\dagger \equiv {\cal H}_{CF}$, then we have
\begin{equation}
{\cal H}_{ex} = {\cal H}_{CF}\otimes{\cal H}_{CS}
\end{equation}
The subspace ${\cal H}_{phy} = {\cal H}_{el}$ satisfies the
constraint,
\begin{equation}
\label{cons}
( \frac{1}{4\pi k} \epsilon_{ij}\partial_i a_j 
-\Psi^{\dagger}(x) \Psi(x) )|\Psi_{phy}>=0
\end{equation}
The contraints in equation (\ref{cons}) are the generators of gauge
transformations and we have,
\begin{eqnarray}
U(\Omega) \Psi(x) U^\dagger (\Omega) = e^{(i \Omega(x))}\Psi(x)\\
U(\Omega) {\vec a}(x) U^\dagger(\Omega) = {\vec a}(x) - {\vec\nabla}\Omega
\end{eqnarray}
where, $ U(\Omega) = e ^{i\int ( \frac{1}{4\pi k} \epsilon_{ij}\partial_i a_j  
- \rho(x) ) \Omega(x)}$ is an element of the gauge group. Thus the physical 
subspace is the gauge invariant subspace of the composite fermion-Chern-Simons
theory.

The electronic observables are hence mapped on to gauge invariant
operators. The ones of interest to us can be explicitly computed to be, 
\begin{eqnarray}
\label{rho}
\rho (x) &=&  \Psi^{\dagger}(x)\Psi(x)\\
\label{jel}
{\vec J}(x) &=& \frac{1}{2i} (\Psi^{\dagger}{\vec\nabla}\Psi-
{\vec\nabla}\Psi^\dagger\Psi) -({\vec A}-{\vec a})\rho(x)\\
\label{cfham0}
H_0  &=& \int d^2 x \Psi^{\dagger} (x) ( {\vec p} - {\vec A} + {\vec a})^2
\Psi (x)\\
\label{cfhamint}
H_{int}&=& \frac{1}{2}\int  U(x-y) \Psi^{\dagger}(x)\Psi^{\dagger}(y) 
\Psi(y)\Psi (x) d^2x d^2y
\end{eqnarray}

\section{Mean field states}
\label{mfs}

\subsection{Physical States}

As seen above, the physical states are the gauge invariant states. We can
construct a projection operator, $P_G$ that projects out the gauge invariant
part of any state.
\begin{equation}
  P_G \equiv \int {\cal D}[\Omega]e^{i\int \frac{1}{4\pi k} 
(\epsilon_{ij}\partial_i a_j -\rho)\Omega}
\end{equation}
$P_G$ basically projects out the singlets of the gauge group. It is easy to
show that,
\begin{equation}
( \frac{1}{4\pi k} \epsilon_{ij}\partial_i a_j 
-\Psi^{\dagger}(x) \Psi(x) )P_G|\Psi>=0
\end{equation}
For example, the empty state of the electron theory is  given by,
\begin{equation}
	|0>_e  = P_G |0>_{CF}\otimes|0>_{CS}
\end{equation}
where,
\begin{equation}
  \Psi (x)|0>_{CF} = 0 = a(x)|0>_{CS}
\end{equation}
  
\subsection{Smoothening the flux: Coherent states} 
 
The mean field theory requires states with uniform Chern-Simons magnetic
fields. A convenient way to construct such states is by constructing a
coherent state basis for the Chern-Simons sector. The details of this
construction can be found in reference \cite{sbdrs}, we outline it here.
  
The displacement operator $ D(\alpha)$, is defined as,
\begin{equation}
D(\alpha) =  e^{\frac{i}{4\pi k} \int \epsilon_{ij} \alpha_i a_j} d^2 x
\end{equation}
The coherent states are obtained by the action of $D(\alpha)$ on the
fluxless state, $|0>_{CS}$ where $a(x)|0>_{CS}=0$,
\begin{equation}
|\alpha> = D(\alpha)|0>_{CS}
\end{equation}
The states $|\alpha>$ satisfy the standard coherent state properties
\cite{sbdrs} including,
\begin{equation}
<\alpha|a(x)|\alpha> = \alpha(x)
\end{equation}
Thus they  are minimum uncertainty wavepackets peaked around
$a(x)=\alpha(x)$. 

To get some more feel for these states, consider the state $|\bar 
\alpha>$ corresponding to a constant magnetic field $b = \nabla 
\times {\vec {\bar \alpha}}$ and compute its overlap with the  
N- vortex state  defined as,
\begin{equation}
|x_1..x_N> = \prod_{m=1}^N e^{{i2k \Omega(x_m)}}P_G |0>_{CS}
\end{equation}
The overlap, which is the amplitude of finding the system in the N-vortex 
state, if it is in the constant field state, can be computed exactly 
using the techniques developed in reference \cite{sbdrs} and is found to be,
\begin{equation}
< x_1..x_N| {\bar\alpha}> = \prod_{n>m} {|x_n -x_m|}^{2k} 
e^{{-\frac{b}{2}\sum_{m=1}^N |x_m|^2}} 
\end{equation}
Thus we see how the constant field state can give rise to the modulus of
the Laughlin factor in the wavefunctions.

\subsection{The Jain states}

We are now ready to construct the gauge invariant mean field states of the 
composite fermion - Chern-Simons theory. Namely, take the direct product
of a composite fermion state with $p$ filled Landau levels in a magnetic field
$B^*=\frac{1}{2kp+1}$, which we denote by $|p,k>_{CF}$ and a constant field 
Chern-Simons state with magnetic field, $b= B-B^*= \frac{2kp}{2kp+1}$, 
denoted by $|p,k>_{CF}$ and gauge project it. We then obtain a state with 
the filling factor $\nu = \frac{p}{2kp+1}$. 
\begin{equation}
\label{js}
|p,k> = P_G |p,k>_{CF}\otimes|p,k>_{CS}
\end{equation}

The wave function of the above state is the overlap with the $N$
electron state, 
\begin{equation}
|\{x_n\}>_{el}=\prod_{n=1}^N C^\dagger(x_n)|0>_{el}
\end{equation}
The overlap can be exactly calculated as shown in reference \cite{sbdrs}
to get,
\begin{equation}
_{el}<\{x_m\}|p,k> = \Phi_{(p,k)}(\{x_m\}) 
\prod_{n>m} (x_n -x_m)^{2k} 
e^{-(\frac{\delta b}{2} \sum_{n=1}^N |x_n|^2 )}
\end{equation}
Where $\Phi_{(p,k)}(\{x_n\})$ is the Slater determinant of the first $p$
Landau levels in a magnetic field $B^*=\frac{1}{2kp+1}$
This is the Jain wave function, in which lowest Landau level projection is 
not made. We will therefore refer to the states defined in equation (\ref{js}) 
as the unprojected Jain states. Note that here "unprojected" refers to
the lowest Landau level and not to the gauge projection. 

\section{The current algebra}
\label{ca}

The electron current and density have been written down in terms of the
composite fermion varaibles in equations (\ref{rho}) and (\ref{jel}).
We defining the composite fermion current to be,
\begin{equation}
\label{jcfdef}
{\vec J}^{CF}(x) = \frac{1}{2i} (\Psi^{\dagger}{\vec\nabla}\Psi-
{\vec\nabla}\Psi^\dagger\Psi) +{\vec b}\rho(x)
\end{equation}
${\vec b}(x)$ is the shifted Chern-Simons field,
\begin{equation}
{\vec b}(x) = {\vec a}(x) + {\vec A}(x)-{\vec A}^*(x)
\end{equation}
Where, ${\vec \nabla}\times{\vec A}^*(x)= B^*$.
The commutators between the components of the currents and the density
can be computed. They form the following closed current algebra.
\begin{eqnarray}
\left [\rho(x),\rho(y) \right ] &=& 0\\ 
\left[J_i^{CF}(x),\rho(y)\right]&=&i\rho(x)\frac{\partial}{\partial y^i} 
\delta^2(x-y)\\
\left [J_i^{CF}(x),J_j^{CF}(y)\right] &=& 
i\epsilon_{ij}B^*\rho(x)\delta^2(x-y)
-iJ_i^{CF}(y)\frac{\partial}{\partial x^j} \delta^2(x-y)
+iJ_j^{CF}(x)\frac{\partial}{\partial y^j} \delta^2(x-y)
\end{eqnarray}
In terms of the fourier transforms, the algebra can be written as,
\begin{eqnarray}
\label{curralgq1}
\left [\rho(q),\rho(q^\prime) \right ] &=& 0\\ 
\label{curralgq2}
\left [J_i^{CF}(q),\rho(q^\prime) \right ] &=& q_i^\prime \rho(q+q^\prime)\\ 
\label{curralgq3}
\left [J_i^{CF}(q),J_j^{CF}(q^\prime)\right] &=& 
i\epsilon_{ij}B^*\rho(q+q^\prime)
+q^\prime_iJ_j^{CF}(q+q^\prime)-q_jJ_i^{CF}(q+q^\prime)
\end{eqnarray}
In the hydrodynamic limit, we are interested in the small amplitude,
long-wavelength fluctuations. So we put,
\begin{equation}
\rho(q+q^\prime)={\bar \rho} \delta^2(q+q^\prime) + \Delta \rho
\end{equation}
where ${\bar \rho} = \nu/2\pi$ and $\Delta \rho/{\bar \rho}<< 1$. Thus in 
the hydrodynamic limit we replace $\rho$ by its average value and take
only the leading terms in $q$. This is what we will mean by the term 
"hydrodynamic limit" from now. The algebra then simplifies considerably
to give,
\begin{eqnarray}
\left [\rho(q),\rho(q^\prime) \right ] &=& 0\\ 
\left [J_i^{CF}(q),\rho(q^\prime) \right ] &=& q_i^\prime{\bar \rho}
\delta^2(q+q^\prime)\\ 
\left [J_i^{CF}(q),J_j^{CF}(q^\prime)\right] &=& 
i\epsilon_{ij}B^*{\bar \rho}\delta^2(q+q^\prime)
\end{eqnarray}
If we define, $J^{CF}(x) = (J^{CF}_1(q)+iJ^{CF}_2(q^\prime))/\sqrt{2B^*{\bar
\rho}}$, then we have the composite fermion current components satisfying the
oscillator algebra,
\begin{equation}
\left[ J^{CF}(q),J^{CF\dagger}(q^\prime)\right]=\delta^2(q-q^\prime)
\end{equation}
In the hydrodynamic limit, the currents behave like the Chern-Simons
gauge fields, they satisfy the same commutation relations.

\section{The hamiltonian and ground state}
\label{th}

\subsection{Current-current form of the hamiltonian}

We will now show that the Hamiltonian can be expressed completely in
terms of the currents and densities in the hydrodynamic limit. The
interaction part of course is already expressed in terms of the
charge densities. So we focus on the non-interacting part, $H_0$, which
can be written as,
\begin{equation}
H_0 = \int~d^2x \frac{1}{2}(\nabla \Psi.\nabla \Psi + 2{\vec J}^{CF}.{\vec b} 
+ \rho{\vec b}.{\vec b} )
\end{equation}
It is straightforward to compute the commutator of the first term in the
above equation with the composite fermion current. In the hydrodynamic
limit, we get,
\begin{eqnarray}
\left[ \int_y \frac{1}{2}(\nabla \Psi.\nabla \Psi), J_i^{CF}(x) \right] 
&=& iB^*\epsilon_{ij}J_j^{CF}(x) \\
&=& \left[ \int_y \frac{1}{2{\bar \rho}}{\vec J}^{CF}(y).{\vec
J}^{CF}(y), J_i^{CF}(x) \right]
\end{eqnarray}
Similarly, we can also show in the hydrodynamic limit that,
\begin{equation}
\left[ \int_y \frac{1}{2}(\nabla \Psi.\nabla \Psi), \rho(x) \right] 
= \left[ \int_y \frac{1}{2{\bar \rho}}{\vec J}^{CF}(y).{\vec J}^{CF}(y), 
\rho(x) \right]
\end{equation}

Therefore, in the hydrodynamic limit, for the dynamics of the charge and
current densities we can make the following replacement in the hamiltonian,
\begin{equation}
\int_y \frac{1}{2}(\nabla \Psi.\nabla \Psi) \rightarrow
\int_y \frac{1}{2{\bar \rho}}{\vec J}^{CF}(y).{\vec J}^{CF}(y)
\end{equation}
The hamiltonian can then be written as,
\begin{equation}
H_0=\int~d^2x \frac{1}{2{\bar \rho}}
({\vec J}^{CF}+{\bar \rho}{\vec b}).({\vec J}^{CF}+{\bar \rho}{\vec b})
\end{equation}
>From equation (\ref{jel}) we see that $({\vec J}^{CF}+{\bar \rho}{\vec
b})$ is the electronic current density with $\rho$ replaced by ${\bar
\rho}$. Thus the hamiltonian has been expressed in the current-current
form as in the case of the other systems.
Defining $b(x)=(b_1(x)+ib_2(x))/\sqrt{8\pi k}$, we can rewrite the
hamiltonian upto a constant as,
\begin{eqnarray}
\label{hamcurr1}
H_0 &=& \int~d^2x 
(cos\theta J^{CF\dagger}+sin\theta b^\dagger)
(cos\theta J^{CF}+sin\theta b)\\
\label{hamcurr2}
&\equiv& \int~d^2x J^\dagger(x)J(x) 
\end{eqnarray}
where, $J \equiv (cos\theta J^{CF}+sin\theta b),~cos\theta \equiv
\sqrt{1/(2kp+1)}~{\rm and}~sin\theta \equiv \sqrt{2kp/(2kp+1)}$.

\subsection{The ground state}

The ground states of $H_0$ (lowest Landau level states) would be those that 
satisfy, 
\begin{equation}
\label{jq0}
J(q)|GS> = 0
\end{equation}
It can be shown that,
\begin{equation}
\label{jcfnl}
\left[ J^{CF}(q), \Psi_{nl} \right] = \sqrt{n}\Psi_{n-1,l+1} + o(q)
\end{equation}
It then follows that, to leading order in $q$,
\begin{equation}
\label{jcfpko}
J^{CF}(q)|p,k>_{CF}=0
\end{equation}
since we also have,
\begin{equation}
b(q)|p,k>_{CS}=0
\end{equation}
we find that the unprojected Jain states are ground states of $H_0$.
All zero energy eigenstates of $H_0$ should be lowest Landau
level states, whereas the unprojected Jain states are not. However, we
should remember that our results are valid only to leading order in $q$.
So these results imply that the non-lowest Landau level component of
the unprojected Jain states does not contribute to matrix elements of
the charge and current density operators to leading order in $q$. This
conclusion will be further bolstered by the consistency of the results
in the next section.

\section{Collective excitations}
\label{ce}

\subsection{Inter-Landau level excitations}

>From equations (\ref{hamcurr1},\ref{hamcurr2}), it follows that the
states,
\begin{equation}
|q;p,k> \equiv J^\dagger(q)|p,k>
\end{equation}
are eigenstates of the hamiltonian with energy = 1 (in our units, $\hbar
\omega_c$). These are therefore,  magnetoplasmons corresponding to inter-Landau
level particle-hole pairs. This is also indicated by equation
(\ref{jcfnl}).

\subsection{Lowest Landau level excitations}

Lowest Landau level excitations will be created by operators that
commute with $H_0$. The following combination of $J^{CF}$ and $b$ that
is "orthogonal" to $J$ does just that.
\begin{equation}
{\tilde J}(x) \equiv \kappa (-sin\theta J^{CF}(x) + cos\theta b(x))
\end{equation}
Using equations (\ref{jel}), (\ref{jcfdef}) and choosing $\kappa$
appropriately, we have,
\begin{equation}
\label{tilj}
{\vec {\tilde J}}(x)= \frac{1}{4\pi k}{\vec b}(x)-{\vec J}(x)  
\end{equation}
${\vec {\tilde J}}(x)$ is not gauge invariant and thus is not a physical
operator. But ${\vec\nabla}\times {\vec{\tilde J}}$ is gauge invariant 
and hence a physical operator. In the physical subspace, we can replace
$\frac{1}{4\pi k} {\vec \nabla}\times{\vec b}$ by $\Delta \rho$. We then
have, 
\begin{equation}
\label{rhoj}
{\vec \nabla} \times {\vec{\tilde J}} = \Delta \rho - 
{\vec\nabla}\times {\vec J}
\end{equation}
Thus we have written ${\vec\nabla}\times {\vec{\tilde J}}$ in terms of
the electronic charge and current densities. To further, discover the
meaning of this quantity, we compute and find the commutators at two
different momenta to be,
\begin{equation}
\left[ i{\vec q}\times{\vec {\tilde J}}(q),
i{\vec q^\prime}\times{\vec {\tilde J}}(q^\prime)\right] = 
i{\vec q}\times{\vec q^\prime}~
(({\vec q}+{\vec q^\prime})\times{\vec{\tilde J}}(q+q^\prime))
\end{equation}
Note, that to get this result, one must use the exact form of the current 
algebra as in equations (\ref{curralgq1}), (\ref{curralgq2}) and 
(\ref{curralgq3}) and use the hydrodynamic approximation
only {\em after} the commutatotrs are computed. Thus,
${\vec\nabla}\times {\vec{\tilde J}}$ obeys the GMP algebra \cite{gmp} to
leading order in $q$. It is therefore natural to identify it with the
charge density operator, projected to the lowest Landau level. This can
also be directly derived from equation (\ref{rhoj}). To do so, consider
first the single particle charge and current density operators.
\begin{eqnarray}
\label{rhospx}
{\hat \rho}(x) &=& \delta^2({\hat x}-x)\\
\label{jspx}
{\hat j}_i(x)  &=& \frac{1}{2}({\hat \Pi}_i{\hat \rho}(x)+
{\hat \rho}(x){\hat \Pi}_i)
\end{eqnarray}
where, ${\hat \Pi}_i = {\hat p}_i - A_i({\hat x})$. Taking fourier
transforms,
\begin{eqnarray}
\label{rhospq}
{\hat \rho}(q) &=& e^{i{\vec q}.{\vec {\hat x}}}\\
\label{jspq}
{\hat j}_i(q)  &=& \frac{1}{2}({\hat \Pi}_i{\hat \rho}(q)+
{\hat \rho}(q){\hat \Pi}_i)
\end{eqnarray}
The lowest Landau level projected density operator is obtained by expressing $\hat x_i$
in terms of $\hat \Pi_i$ and $\hat {\bar \Pi}_i$. Where, ${\hat {\bar \Pi}}_i = 
{\hat p}_i + A_i({\hat x})$. 
\begin{equation}
\label{prhoq}
{\hat {\tilde \rho}}(q) = e^{i{\vec q}\times {\hat{ \vec {\bar \Pi}}}}
\end{equation}
Expanding equations
(\ref{rhospq}), (\ref{jspq}) and (\ref{prhoq}) to leading order in $q$
gives us,
\begin{equation}
{\hat {\tilde \rho}}(q) = {\hat \rho}(q) + 
i{\vec q}\times{\vec{\hat j}}(q)
\end{equation}
Since we have,
\begin{eqnarray}
\rho(q) &=& \int_{x_1,x_2}\Psi^\dagger(x_1)<x_1|{\hat \rho}(q)|x_2>
\Psi(x_2)\\
J_i(q) &=& \int_{x_1,x_2}\Psi^\dagger(x_1)<x_1|{\hat j}_i(q)|x_2>
\Psi(x_2)\\
{\tilde\rho}(q) &=& \int_{x_1,x_2}\Psi^\dagger(x_1)<x_1|
{\hat{\tilde \rho}}(q)|x_2> \Psi(x_2)\\
\end{eqnarray}
it follows that,
\begin{equation}
{\vec \nabla} \times {\vec{\tilde J}} = \Delta {\tilde\rho}  
\end{equation}
It is therefore no accident that it satisfies the GMP algebra to leading
order in $q$. It will therefore, create the lowest Landau level
collective excitations.

Note however, that if we had taken the curl of equation (\ref{tilj}) and
had computed the commutators without using the constraint, then the GMP
algebra would not have been obeyed in general but only in the physical
subspace. Thus if in an approximation scheme, the constraint is not
imposed exactly, as in reference \cite{gmrs1}, it is better to use the
expression in equation (\ref{tilj}) which is what is effectively done in
reference \cite{gmrs1}. However, if the constraint is imposed exactly,
then of course either expression could be used.

\section{Summary}
\label{sum}

As the first step in an attempt to develop a systematic analytic
long-wavelength approximation for FQHE syatems, we have studied the
current algebra of the system in the hydrodynamic limit. Namely, in the 
limit of small amplitude, long-wavelength fluctuations. We develop a 
fully field theoretic Hamiltonian formalism for the Composite fermion theory
following the ideas in reference \cite{gmrs1}. In the hydrodynamical limit,
we show that the components of the current density satisfy very simple 
commutation relations, the same as those satisfied by the components of 
the Chern-Simons fields. The Hamiltonian can be expressed in the
current-current form and the unprojected Jain states are shown to be
ground states. Operators that create inter-Landau level and lowest
Landau level collective excitations are naturally identified using the
current algebra. We show that the operator creating lowest Landau level
collective excitations is precisely the projected charge density
operator and that it satisfies the GMP algebra to leading order in $q$.

\acknowledgements

We thank Sankalpa Ghosh for many useful discussions. M. Sivakumar thanks
DST for support.

\end{document}